\documentclass[prl,showpacs,twocolumn,superscriptaddress,floatfix,10pt,footinbib]{revtex4-2}

\usepackage{setspace} 
\usepackage[utf8]{inputenc}
\usepackage{float}
\usepackage{amsmath, amssymb, amsfonts, bm, bbm}
\usepackage{graphicx}
\usepackage{multirow}
\usepackage{booktabs}
\usepackage[usenames,dvipsnames]{color}
\allowdisplaybreaks
\usepackage{xcolor}
\definecolor{Tiffany}{rgb}{0, 0.7, 0.7}
\definecolor{DarkGreen}{rgb}{0, 0.4, 0}
\usepackage[colorlinks=true,
breaklinks=true,
urlcolor=magenta,
linkcolor=magenta,
citecolor=Tiffany]{hyperref}

\usepackage{orcidlink}

\newcommand{\be}{\begin{equation}}
\newcommand{\ee}{\end{equation}}
\newcommand{\ba}{\begin{align}}
\newcommand{\ea}{\end{align}}
\newcommand{\nt}{\notag\\}

\begin{document}

\title{$H$ dibaryon and its cousins from SU(6)-constrained baryon-baryon interaction}

\begin{abstract}

We constrain the $S$-wave baryon-baryon interaction using SU(6) symmetry within a nonrelativistic effective field theory. The most general leading-order Lagrangian contains two independent parameters, which we determine using physical $NN$ and lattice QCD $\Omega\Omega$ scattering lengths. This framework allows for parameter-free predictions in the strangeness $S=-2$ sector relevant to the $H$ dibaryon. Solving the coupled-channel scattering problem, we identify two bound states below the $\Lambda\Lambda$ threshold, one deeply bound and one shallow, along with resonances near the $\Sigma\Sigma$ and $\Sigma^*\Sigma^*$ thresholds. We demonstrate that these poles result in distinct enhancements in $\Lambda\Lambda$ invariant mass distributions, suggesting that the $H$ dibaryon exists as a multichannel bound state and providing clear signatures for experimental verification.

\end{abstract}

\newcommand{\ucas}{\affiliation{School of Physical Sciences, University of Chinese Academy of Sciences, Beijing 100049, China}}
\newcommand{\itp}{\affiliation{Institute of Theoretical Physics, Chinese Academy of Sciences, Beijing 100190, China}}
\newcommand{\scnt}{\affiliation{Southern Center for Nuclear-Science Theory (SCNT), Institute of Modern Physics,\\ Chinese Academy of Sciences, Huizhou 516000, China}}

\author{Tao-Ran Hu\orcidlink{0009-0003-9720-0171}}\email{hutaoran21@mails.ucas.ac.cn}\ucas\itp
\author{Feng-Kun Guo\orcidlink{0000-0002-2919-2064}}\email{fkguo@itp.ac.cn}\itp\ucas\scnt

\maketitle

Understanding the dynamics of baryon-baryon interactions remains a central problem in low-energy quantum chromodynamics (QCD), with particular relevance to the possible existence of exotic multibaryon states~\cite{Dyson:1964xwa, Locher:1985nu, Epelbaum:2008ga, Clement:2016vnl, Gal:2016boi}. Among these, the strangeness $S=-2$ sector has attracted sustained attention following the long-standing prediction of the $H$ dibaryon~\cite{Jaffe:1976yi}. Its nature---whether deeply bound~\cite{Jaffe:1976yi, Balachandran:1985fb, Yost:1985mj}, weakly bound near threshold~\cite{Straub:1988mz, Kodama:1994np, Nakamoto:1997gh, Shen:2000qs, NPLQCD:2010ocs, Inoue:2010es, Beane:2011zpa, NPLQCD:2011naw, Haidenbauer:2011ah, Haidenbauer:2011za, Inoue:2011ai, Francis:2018qch, Green:2021qol}, or virtual or resonant (quasi-bound)~\cite{Oka:1983ku, Beane:2011zpa, Shanahan:2011su, Haidenbauer:2011ah, Haidenbauer:2011za, Inoue:2011ai, Haidenbauer:2015zqb, Yamaguchi:2016kxa, HALQCD:2019wsz}---is still under active debate. Despite extensive theoretical and experimental efforts, the existence and properties of the $H$ dibaryon remain unsettled, and no conclusive evidence for the $H$ dibaryon has been observed so far, even after dedicated searches in hypernuclear decays, hadron-induced reactions, and high-energy processes~\cite{Iijima:1992pp, BNLE836:1997dwi, E885:2000ped, Takahashi:2001nm, Yoon:2007aq, Belle:2013sba}\footnote{A possibility of finding the $H$ dibaryon in radiative capture reaction $\Lambda\Lambda\to\gamma H$ was proposed in Ref.~\cite{Hikota:2015dba}.}. Yet, the observation of the $_{\Lambda\Lambda}^{~~6}$He hypernucleus has been reported~\cite{Takahashi:2001nm}, which is often interpreted as constraining the lower limit of the $H$ dibaryon mass to around 2223.7~MeV at a 90\% confidence level; however, recent work~\cite{Gal:2024nbr} argues that this constraint alone may still not exclude a deeply bound $H$ dibaryon. The structure of light $S=-2$ hypernuclei and its implications for hyperon-hyperon interactions have been extensively studied; see, e.g., Refs.~\cite{Hiyama:2004be, Contessi:2018qnz, Hiyama:2018lgs, Contessi:2019csf, Le:2021wwz, Le:2021gxa, Miwa:2025adw}.

An appealing organizing principle for baryon-baryon interactions is the approximate spin-flavor symmetry of QCD at low energies. In the large-$N_c$ limit, this symmetry is promoted to SU(6), under which the lowest-lying spin-1/2 octet and spin-3/2 decuplet baryons are unified into a single 56-dimensional multiplet~\cite{Kaplan:1995yg}. When combined with a nonrelativistic effective field theory (NREFT) description, SU(6) symmetry imposes strong constraints on the structure of short-range baryon-baryon interactions, drastically reducing the number of independent low-energy constants (LECs). As a consequence, the leading-order (LO) $S$-wave baryon-baryon interaction can be characterized by a minimal set of parameters, enabling correlated predictions across a wide range of scattering channels. Reference~\cite{Oka:2023hdc} further explored the role of SU(6) symmetry in constraining baryon-baryon interactions, highlighting the importance of the Pauli principle in short-range dynamics. Recently, the consequences of SU(6) symmetry for $NN$, $\Delta\Delta$, and $\Omega\Omega$ scattering were discussed in Ref.~\cite{Richardson:2024zln}, while Refs.~\cite{Vonk:2024lce, Bubpatate:2025eux} analyzed the $1/N_c$ expansion of baryon-baryon potentials in SU(3) chiral effective field theory.

In this work, we investigate the implications of SU(6) symmetry for $S$-wave baryon-baryon scattering in the strangeness $S=-2$ sector. Our analysis is based on an NREFT in which the groundstate SU(3) octet and decuplet baryons are described by a fully symmetric three-index field $\Psi_{\alpha\beta\gamma}$ transforming under the 56-dimensional representation of SU(6). At LO, the most general SU(6)-invariant Lagrangian governing $S$-wave baryon-baryon interactions contains only two independent contact terms, given by~\cite{Kaplan:1995yg}
\begin{align}\label{eq:su6Lag}
\mathcal{L}=&-\tilde{a}~\Psi^\dagger_{\alpha\beta\gamma}\Psi_{\alpha\beta\gamma}\Psi^\dagger_{\mu\nu\rho}\Psi_{\mu\nu\rho}\nt
&-\tilde{b}~\Psi^\dagger_{\alpha\beta\gamma}\Psi_{\alpha\beta\rho}\Psi^\dagger_{\mu\nu\rho}\Psi_{\mu\nu\gamma}\,.
\end{align}
The number of LECs ($\tilde{a},\tilde{b}$) is consistent with that of the antisymmetric SU(6) irreducible representations (irreps):
\begin{equation}
\bm{56} \otimes \bm{56}=\underbrace{\bm{490} \oplus \bm{1050}}_{\rm antisymmetric}\oplus \underbrace{\bm{462}\oplus \bm{1134}}_{\rm symmetric}\,.
\end{equation}
Since baryons obey the Pauli principle, only the antisymmetric combinations are physically relevant. One can construct two linear combinations of the LECs corresponding to these two SU(6) irreps: $\bm{490}\sim2( \tilde{a}-\tilde{b}/3 )$ and $\bm{1050}\sim2( \tilde{a}+\tilde{b}/3 )$, where $\tilde{a}$ and $\tilde{b}$ are the only free parameters governing the baryon-baryon interaction at LO in the SU(6) framework. Remarkably, only two inputs are required to determine these parameters. Once $\tilde{a}$ and $\tilde{b}$ are fixed, the complete baryon-baryon spectrum can be predicted. This simplicity renders the framework highly predictive for the pattern of the dibaryon spectrum and amenable to experimental tests across reactions with various baryon-baryon final states.

A particularly interesting sector of the baryon-baryon spectrum is the one with angular momentum $J=0$, charge $Q=0$, and strangeness $S=-2$ (hence hypercharge $Y=0$ and isospin third component $I_3=0$), which is directly relevant to the study of the $H$ dibaryon. In the particle basis, this sector comprises ten channels, labeled as follows: (1) $\Lambda \Lambda$, (2) $\Lambda \Sigma ^0$, (3) $\Sigma ^0\Sigma ^0$, (4) $\Sigma ^+\Sigma ^-$, (5) $n\Xi ^0$, (6) $p\Xi ^-$, (7) $\Delta ^0\Xi ^{*0}$, (8) $\Delta ^+\Xi ^{*-}$, (9) $\Sigma ^{*0}\Sigma ^{*0}$, (10) $\Sigma ^{*+}\Sigma ^{*-}$. In the isospin $I=0$ sector relevant to the $H$ dibaryon, the number of channels under consideration is reduced to four in the isospin basis. These are labeled as: (1) $\Lambda \Lambda$, (2) $N\Xi^{I=0}$, (3) $\Sigma\Sigma^{I=0}$, (4) $\Sigma ^{*}\Sigma ^{*I=0}$. The four-channel scattering problem will be the primary focus of the following discussion.

Next, we proceed with the formalism for calculating the scattering matrix in this sector. The $T$-matrix, which encodes the scattering information, is defined as:
\begin{equation}
T\left(E\right)=-\left(V^{-1}-G\left(E\right)\right)^{-1},
\end{equation}
with $V=V(\tilde{a},\tilde{b})$ the constant potential matrix read off from the Lagrangian in Eq.~\eqref{eq:su6Lag},
\begin{equation}
\left( \begin{matrix}
	2\tilde{a}&		0&		-\frac{2\sqrt{3}\tilde{b}}{9}&		-\frac{2\sqrt{6}\tilde{b}}{9}\\
	0&		2\left( \tilde{a}-\frac{\tilde{b}}{9} \right)&		-\frac{8\sqrt{3}\tilde{b}}{27}&		\frac{4\sqrt{6}\tilde{b}}{27}\\
	-\frac{2\sqrt{3}\tilde{b}}{9}&		-\frac{8\sqrt{3}\tilde{b}}{27}&		2\left( \tilde{a}+\frac{2\tilde{b}}{27} \right)&		-\frac{2\sqrt{2}\tilde{b}}{27}\\
	-\frac{2\sqrt{6}\tilde{b}}{9}&		\frac{4\sqrt{6}\tilde{b}}{27}&		-\frac{2\sqrt{2}\tilde{b}}{27}&		2\left( \tilde{a}+\frac{\tilde{b}}{27} \right)\\
\end{matrix} \right),
\label{eq:V}
\end{equation}
and the Green's function
\begin{equation}
G=\operatorname{diag}\left(G_1,\ldots,G_{4}\right),\quad G_i=-\frac{\mu_i}{2\pi}\left( \Lambda +ip_i \right),
\end{equation}
where $\mu_i$ is the reduced mass of channel $i$, $p_i=\sqrt{2\mu_i(E-M_i)}$ is the magnitude of the center-of-mass momentum in the channel, $M_i$ is the channel threshold, and $\Lambda$ is the ultraviolet cutoff (not to be confused with the $\Lambda$ baryon), also known as the renormalization scale in the power divergence subtraction (PDS) scheme~\cite{Kaplan:1998tg, Kaplan:1998we} to regularize $G(E)$.

The $S$-wave scattering lengths are defined as $a_i=-\lim_{p_i\rightarrow 0} {\mu _i}T_{ii}/({2\pi})$. In the single-channel case, this convention implies that a positive $a_i$ indicates either the presence of a bound state or a repulsive interaction, while a negative $a_i$ corresponds to an attractive interaction with a virtual state pole.

To determine the LECs $\tilde{a}$ and $\tilde{b}$, two inputs are required. We choose the $NN$ and $\Omega\Omega$ scattering lengths. For $NN$ scattering, there are two partial waves corresponding to isospin 0 and 1: $5.4112(15)$~fm in the $^3S_1$ channel and $-23.7148(43)$~fm in the $^1S_0$ channel~\cite{Hackenburg:2006qd}. The $^1S_0$ $\Omega\Omega$ scattering length is taken from a lattice QCD calculation by the HAL QCD Collaboration: $4.6(6)_\text{stat.}(^{+1.2}_{-0.5})_\text{sys.}$~fm~\cite{Gongyo:2017fjb}. It is important to note that $NN$ scattering is coupled to the $\Delta\Delta$ channel; accordingly, we treat this as a two-channel problem. In contrast, $\Omega\Omega$ scattering involves only a single channel, so no additional complications arise.

When solving for the parameters, we employ two different strategies to relate them to the scattering lengths. Strategy~1 uses the SU(6)-averaged mass of the $\bm{56}$ multiplet of SU(6), denoted $m_{\rm SU(6)}$, while strategy~2 uses the physical values for $m_N$ and $m_\Delta$ and the lattice value for $m_\Omega$~\footnote{In Ref.~\cite{Gongyo:2017fjb}, the $\Omega$ mass is $1712$~MeV, about 2\% higher than the physical value of $1672$~MeV.}. After the LECs are fixed, we use the physical masses for all baryons in the Green's function. In this way, both the $\Lambda$ dependence~\footnote{In the strict SU(6) limit with all masses degenerate everywhere, there would be no $\Lambda$ dependence in the $T$-matrix.} and the difference between the strategies reflect SU(6) breaking effects.

Strategy~1 yields~\cite{Richardson:2024zln}
\begin{align}
\tilde{a}&=-\frac{\pi}{m_{\rm SU(6)}}\frac{\left[ 2a_{\Omega \Omega}\Lambda \left( 9a_{NN}-5a_{\Omega \Omega} \right) -9a_{NN}+a_{\Omega \Omega} \right]}{\left( a_{\Omega \Omega}\Lambda -1 \right) \left( 9a_{NN}\Lambda -5a_{\Omega \Omega}\Lambda -4 \right)}\,, \notag\\
\tilde{b}&=-\frac{27\pi}{m_{\rm SU(6)}}\frac{a_{\Omega \Omega}-a_{NN}}{\left( a_{\Omega \Omega}\Lambda -1 \right) \left( 9a_{NN}\Lambda -5a_{\Omega \Omega}\Lambda -4 \right)}\,.
\end{align}
Within the range of cutoff values considered, this strategy consistently yields a value of $\tilde b$ that is numerically much smaller than $\tilde a$, as shown in Table~\ref{tab:4ch}. Returning to the Lagrangian in Eq.~\eqref{eq:su6Lag}, this hierarchy implies that the first term dominates the short-range interaction at LO, and all channels decouple in the limit $\tilde b=0$---see Eq.~\eqref{eq:V}. A pronounced numerical suppression of $\tilde b$ relative to $\tilde a$ has been observed in lattice calculations performed at pion masses of approximately 806~MeV in Ref.~\cite{Wagman:2017tmp} and 450~MeV in Ref.~\cite{NPLQCD:2020lxg}, which also serve as the PDS scale $\Lambda$ in those studies. The limit $\tilde{b} = 0$ corresponds to an emergent SU(56) symmetry, which has been hypothesized to arise from the vanishing entanglement power of the baryon-baryon scattering $S$-matrix~\cite{Beane:2018oxh, Liu:2022grf, Hu:2025lua}; see also Ref.~\cite{Richardson:2024zln} for a discussion of the possible implications of $\tilde{b} = 0$.

In contrast, strategy~2 yields $\tilde{a}$ and $\tilde{b}$ of comparable magnitude within the cutoff range considered (see Table~\ref{tab:4ch}). This observation suggests that the results from strategy~1 should not be taken as strong evidence for $\tilde{b} \approx 0$. Under strategy~2, one has:
\begin{widetext}
\begin{align}
\tilde{a}&=-\frac{\pi}{\mu_{\Omega}}\frac{\mu _{\Omega}a_{\Omega \Omega}\left( a_{\Omega \Omega}\Lambda -1 \right) \left[ \mu _{\Delta}a_{NN}\left( \Lambda -\gamma \right) +\mu _N\left( a_{NN}\Lambda -1 \right) \right] -9\mu _N\mu _{\Delta}a_{\Omega \Omega}^{2}\left( a_{NN}\Lambda -1 \right) \left( \Lambda -\gamma \right) -9\mu _{\Omega}^{2}a_{NN}\left( a_{\Omega \Omega}\Lambda -1 \right) ^2}{2\left( a_{\Omega \Omega}\Lambda -1 \right) \left\{ 9\mu _N\mu _{\Delta}a_{\Omega \Omega}\left( a_{NN}\Lambda -1 \right) \left( \Lambda -\gamma \right) -\mu _{\Omega}\left( a_{\Omega \Omega}\Lambda -1 \right) \left[ 5\mu _{\Delta}a_{NN}\left( \Lambda -\gamma \right) +4\mu _N\left( a_{NN}\Lambda -1 \right) \right] \right\}}\,, \notag\\
\label{eq:bsol}
\tilde{b}&=-\frac{27\pi}{\mu_{\Omega}}\frac{\left[ \mu _Na_{\Omega \Omega}\left( a_{NN}\Lambda -1 \right) -\mu _{\Omega}a_{NN}\left( a_{\Omega \Omega}\Lambda -1 \right) \right] \left[ \mu _{\Delta}a_{\Omega \Omega}\left( \Lambda -\gamma \right) +\mu _{\Omega}\left( a_{\Omega \Omega}\Lambda -1 \right) \right]}{2\left( a_{\Omega \Omega}\Lambda -1 \right) \left\{ 9\mu _N\mu _{\Delta}a_{\Omega \Omega}\left( a_{NN}\Lambda -1 \right) \left( \Lambda -\gamma \right) -\mu _{\Omega}\left( a_{\Omega \Omega}\Lambda -1 \right) \left[ 5\mu _{\Delta}a_{NN}\left( \Lambda -\gamma \right) +4\mu _N\left( a_{NN}\Lambda -1 \right) \right] \right\}}\,,
\end{align}
\end{widetext}
where $\mu_N=m_N/2$, $\mu_\Delta=m_\Delta/2$, $\mu_\Omega=m_\Omega/2$, $\gamma=\sqrt{2\mu_\Delta(2m_\Delta-2m_N)}$.

In summary, four sets of $(\tilde{a},\tilde{b})$ are obtained, corresponding to the use of these two strategies along with the application of two different values for the $NN$ scattering length.

Using these LECs, we predict the baryon-baryon scattering lengths, as shown in Table~\ref{tab:4ch}. Since the $\Lambda\Lambda$ channel has the lowest threshold among the four isoscalar coupled channels with strangeness $S=-2$, scattering lengths in all other channels are complex. The quoted uncertainties combine in quadrature the propagated uncertainty from the $\Omega\Omega$ scattering length and the cutoff dependence, with the former negligible in most cases. Central values correspond to $\Lambda = 1.0$~GeV, while $\Lambda$ is varied from 2.0 to 0.5~GeV to estimate higher-order corrections omitted in the LO NREFT and would vanish if all baryon masses were degenerate also in the Green's function.

\begin{figure}[tb]
    \begin{center}
    \includegraphics[width=\linewidth]{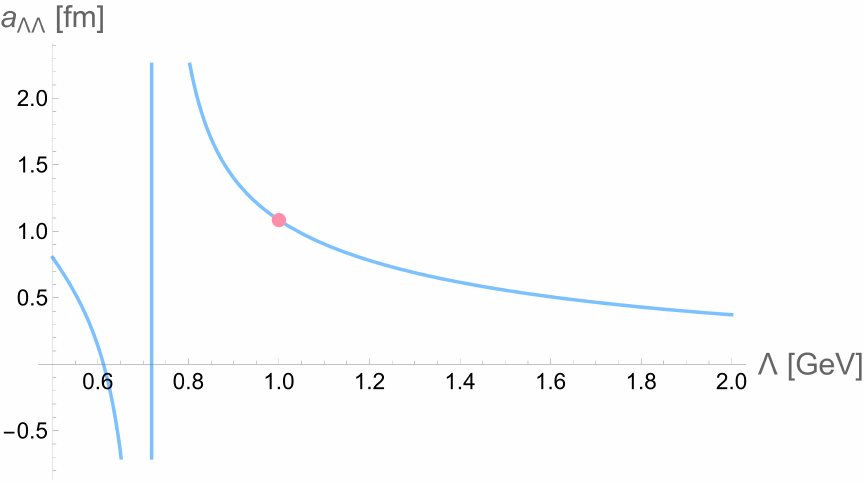}
    \caption{Cutoff dependence of the $\Lambda\Lambda$ scattering length $a_{\Lambda\Lambda}$ in strategy~2 with $a_{NN}$ in $^3S_1$ as input. The dot marks the $a_{\Lambda\Lambda} = 1.1$~fm at cutoff $\Lambda = 1.0$~GeV. At $\Lambda = 0.72$~GeV, $a_{\Lambda\Lambda}$ reaches the unitarity limit.}
    \label{fig:cutoff}
    \end{center}
\end{figure}

Two observations are immediate. First, for all four parameter sets, both the $\Lambda\Lambda$ scattering length and the real part of the $N\Xi$ scattering length are large in magnitude, indicating the presence of poles within a few tens of MeV of the respective thresholds. In particular, the $\Lambda\Lambda$ scattering length in strategy~2 crosses the unitarity limit as the cutoff varies, jumping from positive to negative infinity, as shown in Fig.~\ref{fig:cutoff}.

\begin{table*}[tb]
\centering
\caption{Solved LECs and predicted baryon-baryon scattering lengths for different strategies using $NN$ scattering lengths in the $^3S_1$ and $^1S_0$ channels. The uncertainties include both the propagated uncertainty from the $\Omega\Omega$ input and the cutoff dependence (central values correspond to cutoff $\Lambda=1.0$~GeV). For strategy~1, the imaginary parts of the scattering lengths, as well as their associated uncertainties, are of order $0.01$~fm and are therefore not reported. For strategy~2, some channels cross the unitarity limit as the cutoff is varied, so instead of reporting their uncertainties, we present the range of their values in a piecewise manner where the $\Lambda$ value is in units of GeV; see also Fig.~\ref{fig:cutoff}.}
\begin{tabular}{p{5em}cccc}
\toprule
\multirow{2}{*}{Strategies} & \multicolumn{2}{c}{Strategy~1} & \multicolumn{2}{c}{Strategy~2} \\
\cmidrule(lr){2-3}\cmidrule(lr){4-5}
& \quad with $a_{NN}$ in $^3S_1$\quad & \quad with $a_{NN}$ in $^1S_0$\quad & \quad with $a_{NN}$ in $^3S_1$\quad & \quad with $a_{NN}$ in $^1S_0$\quad \\
\midrule
$\tilde{a}$ [GeV$^{-2}$]
& $-5.0\left(^{+2.5}_{-5.4}\right)$
& $-4.9\left(^{+2.5}_{-5.1}\right)$
& $-6.9\left(^{+4.1}_{-12.3}\right)$
& $-6.7\left(^{+3.8}_{-9.4}\right)$ \\[5pt]
$\tilde{b}$ [GeV$^{-2}$]
& $0.1\left(^{+0.3}_{-0.2}\right)$
& $0.4\left(^{+1.1}_{-0.3}\right)$
& $-9.3\left(^{+6.3}_{-24.2}\right)$
& $-8.5\left(^{+5.6}_{-15.7}\right)$ \\
\midrule
$a_{\Lambda\Lambda}$ [fm]
& $-1.5\left(^{+0.8}_{-2.8}\right)$
& $-1.3\left(^{+0.7}_{-1.7}\right)$
& $1.1\left(^{\Lambda>0.72:~[0.4,+\infty)}_{\Lambda<0.72:~(-\infty,0.8]}\right)$
& $2.2\left(^{\Lambda>0.90:~[0.4,+\infty)}_{\Lambda<0.90:~(-\infty,1.4]}\right)$ \\[5pt]
$a_{N\Xi}^{I=0}$ [fm]
& $-1.3\left(^{+0.7}_{-2.4}\right)$
& $-1.2\left(^{+0.7}_{-1.8}\right)$
& \quad$0.6\left(^{+0.9}_{-0.4}\right)-i0.08\left(^{+0.1}_{-0.1}\right)$\quad
& \quad$0.7\left(^{+1.9}_{-0.5}\right)-i0.12\left(^{+0.2}_{-0.1}\right)$\quad \\[5pt]
$a_{\Sigma\Sigma}^{I=0}$ [fm]
& $-3.1\left(^{+1.9}_{-12.8}\right)$
& $-2.29\left(^{+1.3}_{-3.3}\right)$
& \quad$0.2\left(^{+0.4}_{-0.1}\right)-i0.3\left(^{+0.4}_{-0.2}\right)$\quad
& \quad$0.3\left(^{+0.6}_{-0.1}\right)-i0.3\left(^{+0.5}_{-0.2}\right)$\quad \\[5pt]
$a_{\Sigma^*\Sigma^*}^{I=0}$ [fm]
& $2.4\left(^{+1.0}_{-0.9}\right)$
& $3.0\left(^{+1.8}_{-1.3}\right)$
& \quad$0.5\left(^{+0.6}_{-0.3}\right)-i0.4\left(^{+0.1}_{-0.2}\right)$\quad
& \quad$0.5\left(^{+0.7}_{-0.3}\right)-i0.4\left(^{+0.1}_{-0.2}\right)$\quad \\
\bottomrule
\end{tabular}
\label{tab:4ch}
\end{table*}

Given the large scattering lengths, we now search for poles of the $T$-matrix on the complex energy plane across various Riemann sheets (RSs). For strategy~1 with $a_{NN}$ in $^3S_1$, the four channels are essentially decoupled (as seen from the values of $\tilde a$ and $\tilde b$), allowing separate single-channel analyses. The $\Sigma^*\Sigma^*$ channel exhibits a bound state, while the $\Lambda\Lambda$, $\Sigma\Sigma$, and $N\Xi$ channels each display a virtual state, as implied by the signs of the scattering lengths. Including coupled-channel effects produces negligible shifts of these poles. The results using $a_{NN}$ in $^1S_0$ are analogous. These pole positions are summarized in Table~\ref{tab:ES1}.

\begin{table}[tb]
\centering
\caption{Pole positions obtained using strategy~1 and single-channel analyses. The uncertainties include contributions from the $\Omega\Omega$ scattering length and from the cutoff dependence, estimated by scanning the cutoff $\Lambda$ from 2.0 to 0.5~GeV around the central value $\Lambda=1.0$~GeV. The subscript ``$\pm$'' denotes a pole on the first/second RS, corresponding to a bound/virtual state, respectively. }
\begin{tabular}{p{6em}cc}
\toprule
\multirow{2}{*}{} & \multicolumn{2}{c}{Strategy~1} \\
\cmidrule(lr){2-3}
& with $a_{NN}$ in $^3S_1$ & with $a_{NN}$ in $^1S_0$ \\
\midrule
$E_{\Lambda\Lambda}$ [GeV]
& $2.21\left(^{+0.01}_{-0.07}\right)_-$
& $2.21\left(^{+0.02}_{-0.08}\right)_-$ \\[5pt]
$E_{N\Xi}$ [GeV]
& $2.24\left(^{+0.02}_{-0.09}\right)_-$
& $2.23\left(^{+0.02}_{-0.10}\right)_-$ \\[5pt]
$E_{\Sigma\Sigma}$ [GeV]
& $2.38\left(^{+0.00}_{-0.02}\right)_-$
& $2.38\left(^{+0.01}_{-0.02}\right)_-$ \\[5pt]
$E_{\Sigma^*\Sigma^*}$ [GeV]
& $2.76\left(^{+0.00}_{-0.01}\right)_+$
& $2.77\left(^{+0.00}_{-0.02}\right)_+$ \\
\bottomrule
\end{tabular}
\label{tab:ES1}
\end{table}

For strategy~2 with $a_{NN}$ in $^3S_1$ as input, we find a total of thirty-two poles: two bound states and ten virtual states below the $\Lambda\Lambda$ threshold, along with ten complex-conjugate pairs of resonance poles between the $\Lambda\Lambda$ and $\Sigma^*\Sigma^*$ thresholds. Among these, two bound states and two resonance pairs are dominant, while the remaining twenty-six are shadow poles~\cite{Eden:1964zz}. The results using $a_{NN}$ in $^1S_0$ are analogous. We label the RSs as $(\pm,\pm,\pm,\pm)$, where the signs denote the imaginary parts of the channel momenta; bound states lie on the real axis of the physical sheet $(+,+,+,+)$. The dominant pole positions and their channel couplings are listed in Table~\ref{tab:ES2}.

\begin{table}[tb]
\centering
\caption{Four dominant poles and their coupling strengths to the scattering channels obtained using strategy~2. Here the RSs $(+,+,+,+)$, $(-,+,+,+)$, $(-,-,+,+)$, and $(-,-,-,+)$ are denoted as RS~I, II, III, and IV, respectively. Pole energies and coupling strengths are in units of GeV and GeV$^{-1/2}$, respectively. The uncertainties include contributions from the $\Omega\Omega$ scattering length and from the cutoff dependence, estimated by scanning the cutoff $\Lambda$ from 2.0 to 0.5~GeV around the central value $\Lambda=1.0$~GeV; the former is found to be numerically negligible compared to the latter. The cutoff dependence of the pole positions is also illustrated in Fig.~\ref{fig:trajectory}. Notably, the second pole in the RS~I moves to the RS~II as $\Lambda$ decreases; see its trajectory in Fig.~\ref{fig:trajectory}.}
\begin{tabular}{p{2em}cc}
\toprule
\multirow{2}{*}{} & \multicolumn{2}{c}{Strategy~2} \\
\cmidrule(lr){2-3}
&  with $a_{NN}$ in $^3S_1$ & with $a_{NN}$ in $^1S_0$ \\
\midrule
$E_{\rm I}^{(1)}$
& $2.13\left(^{+0.07}_{-0.11}\right)$
& $2.16\left(^{+0.06}_{-0.11}\right)$ \\[5pt]
$g_{\Lambda\Lambda}$
& $1.74\left(^{+0.06}_{-0.05}\right)$
& $1.71\left(^{+0.07}_{-0.19}\right)$ \\[5pt]
$g_{N\Xi}^{I=0}$
& $1.11\left(^{+0.16}_{-0.50}\right)$
& $1.03\left(^{+0.21}_{-0.63}\right)$ \\[5pt]
$g_{\Sigma\Sigma}^{I=0}$
& $1.91\left(^{+0.51}_{-0.88}\right)$
& $1.78\left(^{+0.58}_{-1.04}\right)$ \\[5pt]
$g_{\Sigma^*\Sigma^*}^{I=0}$
& $0.27\left(^{+0.07}_{-0.01}\right)$
& $0.28\left(^{+0.06}_{-0.04}\right)$ \\
\midrule
$E_{\rm I}^{(2)}$
& $2.22\left(^{+0.03-i0.003_{\rm II}}_{-0.09_{\rm I}}\right)$
& $2.23\left(^{+0.03-i0.003_{\rm II}}_{-0.09_{\rm I}}\right)$ \\[5pt]
$g_{\Lambda\Lambda}$
& $1.04\left(^{+2.84}_{-0.63}\right)$
& $0.78\left(^{+0.71}_{-0.67}\right)$ \\[5pt]
$g_{N\Xi}^{I=0}$
& $1.33\left(^{+8.27}_{-0.40}\right)$
& $1.20\left(^{+1.47}_{-0.97}\right)$ \\[5pt]
$g_{\Sigma\Sigma}^{I=0}$
& $0.80\left(^{+5.84}_{-0.18}\right)$
& $0.78\left(^{+1.15}_{-0.63}\right)$ \\[5pt]
$g_{\Sigma^*\Sigma^*}^{I=0}$
& $0.95\left(^{+3.00}_{-0.59}\right)$
& $0.76\left(^{+1.18}_{-0.64}\right)$ \\
\midrule
$E_{\rm III}$
& $2.47\left(^{+0.19}_{-0.08}\right)-i0.11\left(^{+0.28}_{-0.08}\right)$
& $2.46\left(^{+0.19}_{-0.07}\right)-i0.10\left(^{+0.27}_{-0.12}\right)$ \\[5pt]
$g_{\Lambda\Lambda}$
& $0.82\left(^{+0.01}_{-0.40}\right)$
& $0.82\left(^{+0.01}_{-0.38}\right)$ \\[5pt]
$g_{N\Xi}^{I=0}$
& $1.87\left(^{+1.15}_{-0.87}\right)$
& $1.81\left(^{+1.15}_{-0.96}\right)$ \\[5pt]
$g_{\Sigma\Sigma}^{I=0}$
& $2.30\left(^{+0.38}_{-0.51}\right)$
& $2.27\left(^{+0.38}_{-0.60}\right)$ \\[5pt]
$g_{\Sigma^*\Sigma^*}^{I=0}$
& $0.48\left(^{+1.19}_{-0.34}\right)$
& $0.45\left(^{+1.16}_{-0.34}\right)$ \\
\midrule
$E_{\rm IV}$
& $2.77\left(^{+0.12}_{-0.01}\right)-i0.07\left(^{+0.21}_{-0.05}\right)$
& $2.76\left(^{+0.11}_{-0.01}\right)-i0.06\left(^{+0.21}_{-0.05}\right)$ \\[5pt]
$g_{\Lambda\Lambda}$
& $0.95\left(^{+0.90}_{-0.53}\right)$
& $0.90\left(^{+0.93}_{-0.54}\right)$ \\[5pt]
$g_{N\Xi}^{I=0}$
& $0.64\left(^{+0.57}_{-0.35}\right)$
& $0.61\left(^{+0.59}_{-0.36}\right)$ \\[5pt]
$g_{\Sigma\Sigma}^{I=0}$
& $0.19\left(^{+0.18}_{-0.11}\right)$
& $0.18\left(^{+0.18}_{-0.11}\right)$ \\[5pt]
$g_{\Sigma^*\Sigma^*}^{I=0}$
& $1.96\left(^{+0.59}_{-0.45}\right)$
& $1.92\left(^{+0.63}_{-0.50}\right)$ \\
\bottomrule
\end{tabular}
\label{tab:ES2}
\end{table}

We now analyze the results in more detail, focusing exclusively on strategy~2 with $a_{NN}$ in the $^3S_1$ channel. A salient feature is the presence of strong coupled-channel effects. The $\Lambda\Lambda$ threshold lies at $2231.4$~MeV. The deeper bound state is located tens of MeV below it; within uncertainties, however, the binding energy may be as small as 10~MeV, compatible with the lower bound on the $H$ dibaryon mass reported in Ref.~\cite{Takahashi:2001nm}. This state couples strongly to the $\Lambda\Lambda$, $N\Xi$, and $\Sigma\Sigma$ channels (see the couplings in Table~\ref{tab:ES2}), indicating a pronounced multichannel character. The shallower bound state, which could become a resonance pole on the $(-,+,+,+)$ RS within uncertainties, lies closer to the $\Lambda\Lambda$ threshold and also near the $N\Xi$ threshold at around 2.26~GeV, coupling predominantly to the $N\Xi$ channel. Both poles can be regarded as members of the $S=-2$ $H$ dibaryon family. Notably, Ref.~\cite{Haidenbauer:2011ah} analyzed the $H$ dibaryon within chiral effective field theory and concluded that, if bound, it is more likely to be an $N\Xi$ bound state---a scenario also discussed in quark-model studies~\cite{Oka:1983ku, Nakamoto:1997gh}---with a binding energy of roughly 28~MeV relative to the $N\Xi$ threshold.

The resonance pole on RS $(-,-,+,+)$ couples most strongly to the $\Sigma\Sigma$ channel (see Table~\ref{tab:ES2}), with its imaginary part arising primarily from decays into the lower-lying channels. Notably, the real part of this pole lies above the $\Sigma\Sigma$ threshold. This can be understood as a consequence of the strong $N\Xi$-$\Sigma\Sigma$ coupling, which pushes this pole upward in energy while simultaneously driving the deeply bound state further down (for a general discussion, see Ref.~\cite{Zhang:2024qkg}). The resonance pole on RS $(-,-,-,+)$ can be interpreted as a $\Sigma^*\Sigma^*$ quasi-bound state, to which it couples predominantly. The $\Sigma^*\Sigma^*$ channel couples relatively weakly to the other three channels, as expected from the large mass gap between the spin-$3/2$ and spin-$1/2$ baryons, which leads to a partial decoupling.

\begin{figure*}[tb]
    \centering
        \includegraphics[height=4.cm]{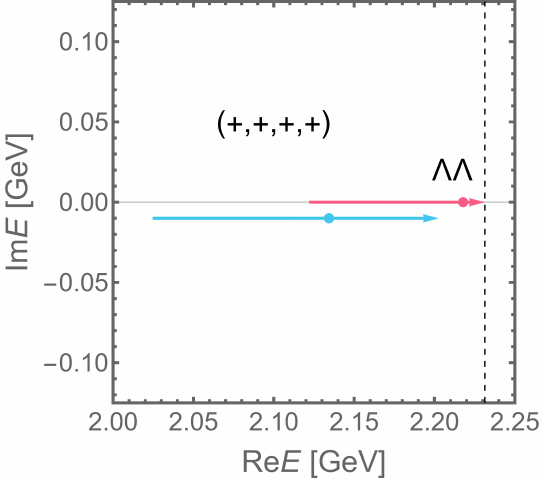}\hfill
        \includegraphics[height=4.cm]{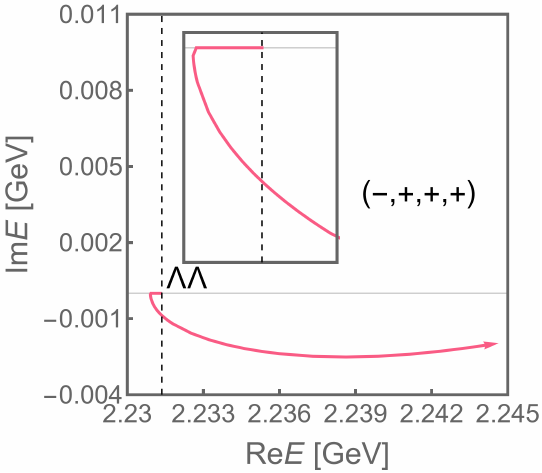}\hfill
        \includegraphics[height=4.cm]{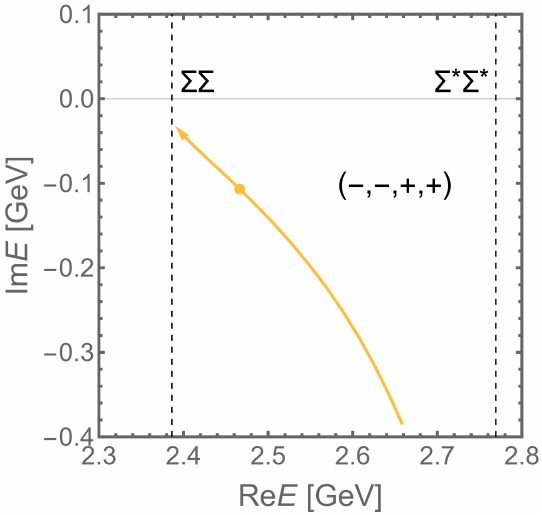}\hfill
        \includegraphics[height=4.cm]{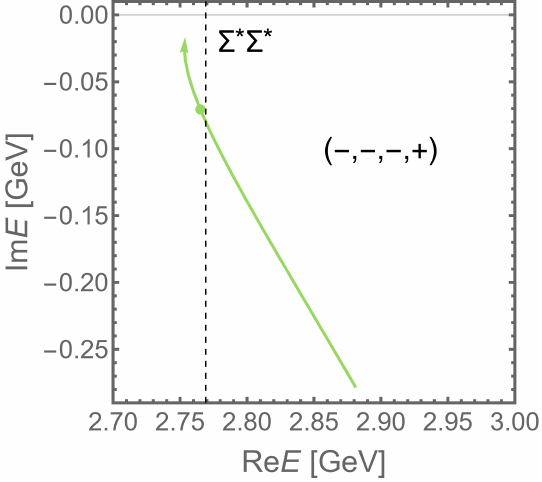}
    \caption{Trajectories of the four dominant poles in strategy~2 with $a_{NN}$ in $^3S_1$ as input, with arrows indicating decreasing the cutoff $\Lambda$ from 2.0~GeV to 0.5~GeV. Highlighted points mark the values at $\Lambda = 1.0$~GeV. On RS $(+,+,+,+)$, the deeper bound state is shifted slightly downward to avoid overlap with the shallower one. }
    \label{fig:trajectory}
\end{figure*}

Next, we examine the cutoff dependence of these four dominant poles by scanning $\Lambda$ from 2.0~GeV down to 0.5~GeV and tracing the pole trajectories on the RSs, as shown in Fig.~\ref{fig:trajectory}. For the two bound state poles on RS $(+,+,+,+)$, the binding becomes progressively shallower as the cutoff decreases. The shallower bound state reaches the $\Lambda\Lambda$ threshold at $\Lambda = 0.72$~GeV and, upon further reduction, moves onto RS $(-,+,+,+)$ as a virtual state. It then shifts along the negative real axis and eventually acquires an imaginary part, becoming a resonance---a clear indication that the attraction weakens with decreasing cutoff. For the two resonance poles on RS $(-,-,+,+)$ and $(-,-,-,+)$, their real parts decrease and both approach the real axis as the cutoff is lowered. This pattern can be understood as follows: decreasing the cutoff reduces the couplings between the lower and higher channel pairs, thereby weakening both the coupling-induced effective attraction in the $\Lambda\Lambda$ and $N\Xi$ channels and the coupling-induced effective repulsion in the $\Sigma\Sigma$ and $\Sigma^*\Sigma^*$ channels.

\begin{figure}[tb]
    \begin{center}
    \includegraphics[width=\linewidth]{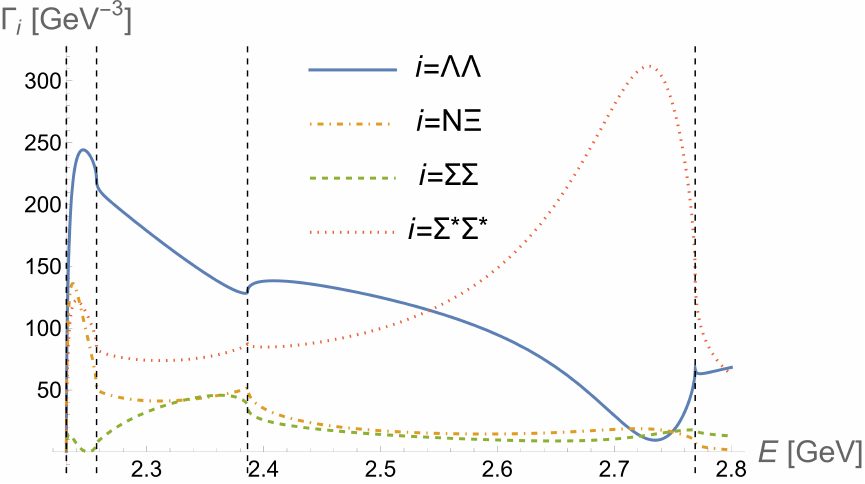}
    \caption{Invariant mass distributions for the processes $\Lambda\Lambda$, $N\Xi$, $\Sigma\Sigma$, and $\Sigma^{*}\Sigma^{*}\to\Lambda\Lambda$ in strategy~2 with $a_{NN}$ in $^3S_1$, evaluated at cutoff $\Lambda = 1.0$~GeV. The vertical dashed lines indicate the four thresholds.}
    \label{fig:ls}
    \end{center}
\end{figure}

Having clarified the analytic structure of the $T$-matrix, we now examine how these poles manifest in the $\Lambda\Lambda$ invariant mass spectrum. The invariant mass distribution for reactions induced via $i \to \Lambda\Lambda$ is proportional to
\begin{equation}
\Gamma_i=p_{\Lambda\Lambda}\left|T_{i\to\Lambda\Lambda}\right|^2,
\end{equation}
and the resulting lineshapes for the four channels are shown in Fig.~\ref{fig:ls}. A pronounced enhancement appears near the $\Lambda\Lambda$ threshold, arising from the lowest two poles. This feature is most prominent in the $\Lambda\Lambda \to \Lambda\Lambda$ channel, where the peak reaches its maximum around $2246$~MeV. In reactions with two $\Lambda$ hyperons in the final state, the $\Lambda\Lambda$ pair can proceed through all possible intermediate states before being observed, i.e., it can be produced via rescattering of other baryon pairs. In general, one expects that the $\Lambda\Lambda$ channel can be more easily produced than other $S=-2$ baryon pairs for several reasons: (1) the $\Lambda\Lambda$ threshold is the lowest; (2) forming an $N\Xi$ pair requires the two strange quarks to coalesce into a single $\Xi$, which demands a small relative momentum and is thus suppressed; (3) heavier hyperons decay into $\Lambda$. In such reactions, one expects to see a peak near the $\Lambda\Lambda$ threshold in the $\Lambda\Lambda$ invariant mass spectrum; as reported in the preliminary result of J-PARC E42 experiment~\cite{Ahn:2025exotic}.

Beyond the bound states, the two resonance poles also leave characteristic imprints on the invariant mass distributions. The lower-lying resonance couples most strongly to $\Sigma\Sigma$ and lies above the $\Sigma\Sigma$ threshold. Consequently, it produces a cusp at the $\Sigma\Sigma$ threshold in the other three lineshapes, manifesting either as a dip~\cite{Dong:2020hxe} or a peak. In the $\Sigma\Sigma\to\Lambda\Lambda$ transition, it generates an asymmetric enhancement that peaks below the $\Sigma\Sigma$ threshold and is strongly distorted and suppressed by the nearby threshold---a generic feature of resonances that are inaccessible on the physical RS in off-diagonal $T$-matrix elements~\cite{Zhang:2024qkg}. By contrast, the higher-lying resonance is unimpeded by the distant lower thresholds and thus appears as a clear peak in $\Gamma_{\Sigma^*\Sigma^*}$, while simultaneously inducing a pronounced dip in $\Gamma_{\Lambda\Lambda}$---a universality feature discussed in Ref.~\cite{Dong:2020hxe}.

In conclusion, we have investigated the implications of SU(6) symmetry for $S$-wave baryon-baryon scattering in the $S=-2$ sector. Using an LO NREFT, the baryon-baryon interaction is constrained to two LECs, $\tilde{a}$ and $\tilde{b}$, enabling a predictive framework testable across multiple scattering channels. Focusing on the $J=0$, $Q=0$, $S=-2$ sector, which is of particular interest for the $H$ dibaryon, we find two bound states, consistent with the $H$ dibaryon manifesting as a multichannel state. There are also two higher poles located at about 2.47 and 2.77~GeV, respectively. In particular, the higher one should appear as a pronounced dip in the $\Lambda\Lambda$ invariant mass distribution of a reaction producing much more $\Lambda\Lambda$ than $\Sigma^*\Sigma^*$ pairs. These results offer new insights into exotic multibaryon dynamics and motivate further experimental searches for the $H$ dibaryon.

\begin{acknowledgements}
We thank Ulf-G. Mei{\ss}ner for a careful reading of the manuscript and helpful comments. This work is supported in part by the National Key R\&D Program of China under Grant No.~2023YFA1606703; by the National Natural Science Foundation of China under Grants No.~12125507, No.~12361141819, and No.~12447101; and by the Chinese Academy of Sciences under Grant No.~YSBR-101.
\end{acknowledgements}

\bibliography{refs}

\end{document}